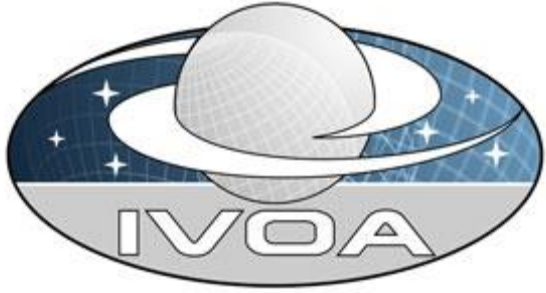

International

Virtual

Observatory

Alliance

# HiPS – Hierarchical Progressive Survey

## Version 1.0
### IVOA Recommendation
### 19th May 2017

**Editor:**
> Pierre Fernique

**Authors:**
> Pierre Fernique, Mark Allen, Thomas Boch, Tom Donaldson, Daniel Durand,
> Ken Ebisawa, Laurent Michel, Jesus Salgado, Felix Stoehr

---

## Abstract

This document presents HiPS, a hierarchical scheme for the description, storage and access of sky survey data. The system is based on hierarchical tiling of sky regions at finer and finer spatial resolution which facilitates a progressive view of a survey, and supports multi-resolution zooming and panning. HiPS uses the HEALPix tessellation of the sky as the basis for the scheme and is implemented as a simple file structure with a direct indexing scheme that leads to practical implementations.

# Status of This Document

This document has been reviewed by IVOA Members and other interested parties, and has been endorsed by the IVOA Executive Committee as an IVOA Recommendation. It is a stable document and may be used as reference material or cited as a normative reference from another document. IVOA's role in making the Recommendation is to draw attention to the specification and to promote its widespread deployment. This enhances the functionality and interoperability inside the Astronomical Community.

*A list of current IVOA Recommendations and other technical documents can be found at http://www.ivoa.net/Documents/.*

# Acknowledgments

We acknowledge support from the Astronomy ESFRI and Research Infrastructure Cluster – ASTERICS project, funded by the European Commission under the Horizon 2020 Programme (GA 653477).

Content

# 1 Introduction

This document describes the Hierarchical Progressive Survey (HiPS) system for access and visualization of astronomical survey data. The aim of the HiPS system is to enable dedicated client/browser tools to access and display an astronomical survey progressively, based on the principle that "*the more you zoom in on a particular area the more details show up*".

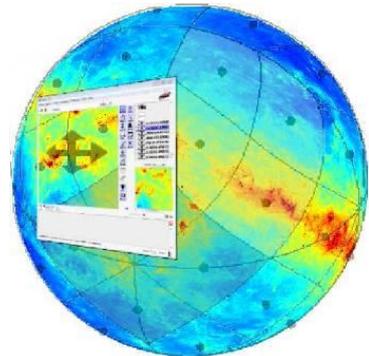

We adopt the term 'astronomical survey' in a broad sense to include: all-sky surveys, collections of pointed observations, catalogues of sources. In principle HiPS can manage any types of data that can be located and displayed on the celestial sphere.

HiPS is designed specifically for astronomical data in that it takes the astrometric and photometric properties of the original data into account, and emphasis is placed on ease of use without the need for special servers or database systems.

The full context and motivation of HiPS is described in Fernique 2015 [1]. HiPS was originally designed by CDS (Centre de Données astronomiques de Strasbourg) in 2010 in the framework of the Aladin project whereupon its popularity has rapidly overtaken the original scope. The emergence of several new and independent HiPS clients and servers has been the main motivation to standardize this method in the IVOA framework. As hundreds of HiPS are already available and in use, this document aims to combine the existing pre-IVOA HiPS conventions and formulate a new IVOA standard for it.

Essentially, HiPS describes a generic method for packaging, storing, querying, and describing astronomical data. There are dependencies on other VO standards, notably MOC [4], the Resource Metadata [2], the IVOA Identifier [3], the ObsCoreDM [8] and VOTable [9]. HIPS is also based on other non-IVOA standards notably HEALPix [5] and FITS [6], as well as the JPEG and PNG image formats.

The figure above illustrates HiPS in the context of the overall IVOA Architecture.

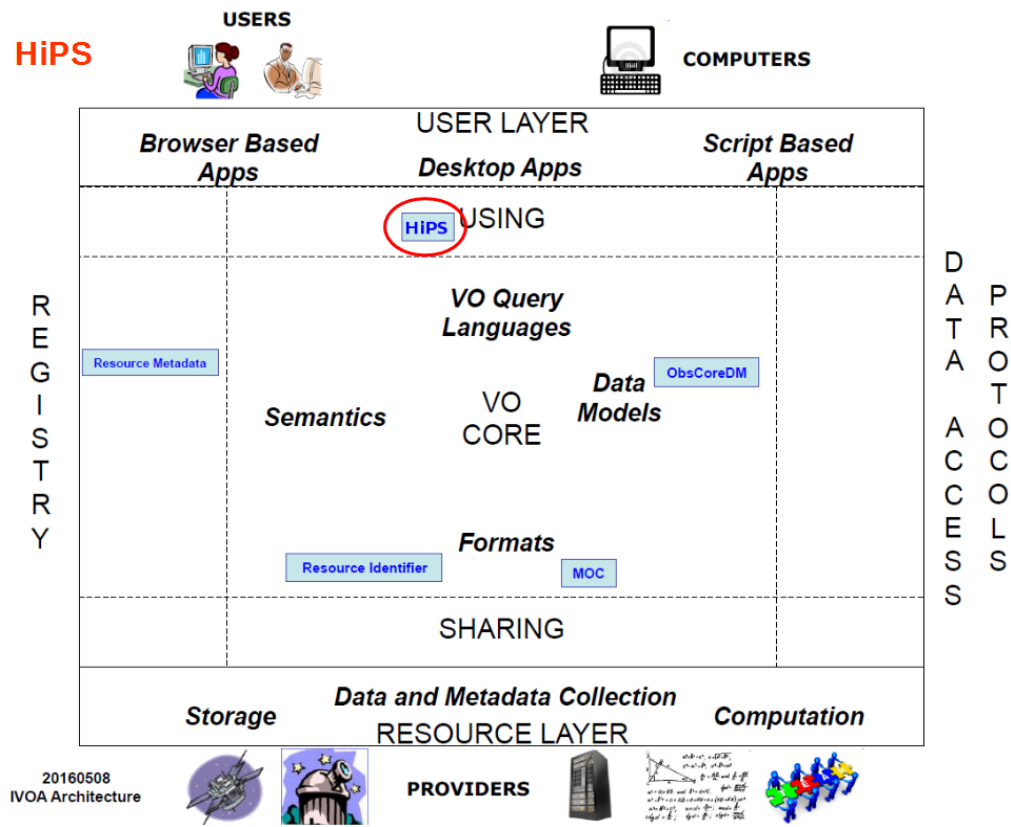

## 2  Usage examples

The most common usage of HiPS is the visualisation of data from large astronomical surveys. HiPS allows one to browse "big data": pan and zoom into each section of the survey data using HiPS clients that access the data over the internet. Only the portion of the data needed for the current user view is streamed from the server to the client. This remote visualisation technique enables the exploration of large data sets from a wide field of view where an entire survey is projected on the whole sky, to a detailed zoomed view at the finest spatial resolution of the images used for the survey.

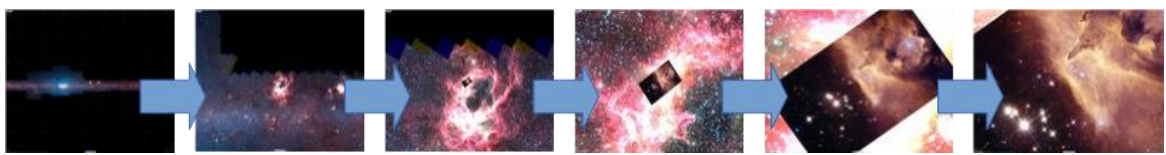

This visualisation technique is not limited to pixel surveys but may also be used for other data types: source catalogues, pixel cubes, etc. For example, the HiPS scheme can be used to describe a multiresolution view of a source catalogue where the selection of catalogue points is a representation of different parameters like brightness. This allows a different view of the catalogue at different zoom level. For example we could display fainter and fainter sources as one zooms into higher orders of the map. As such, HiPS catalogues can provide a hierarchical view of catalogues where the hierarchy can be based on a selected quantitative property for each catalogue.

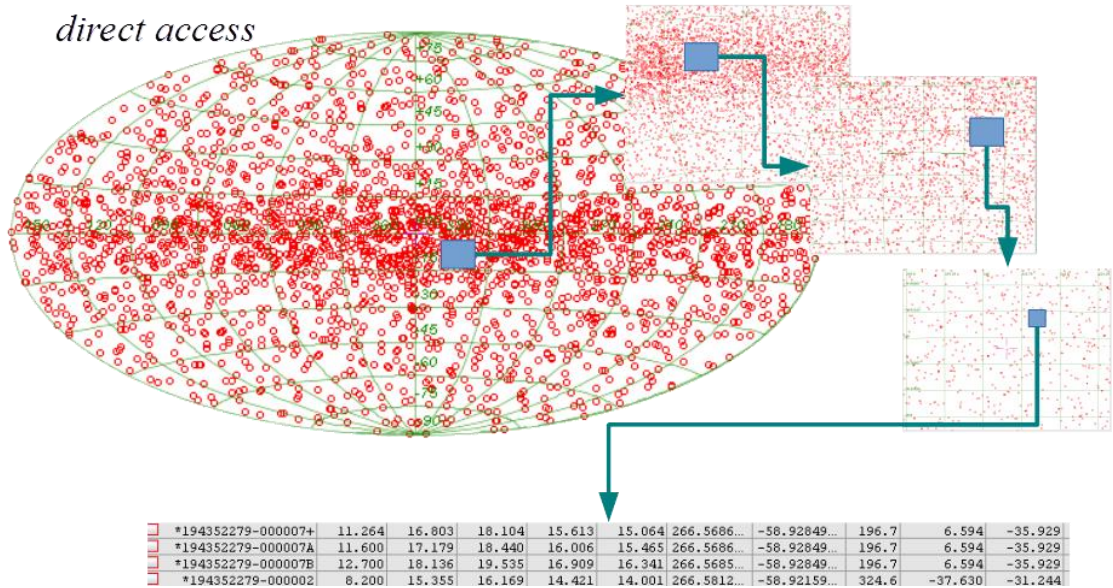

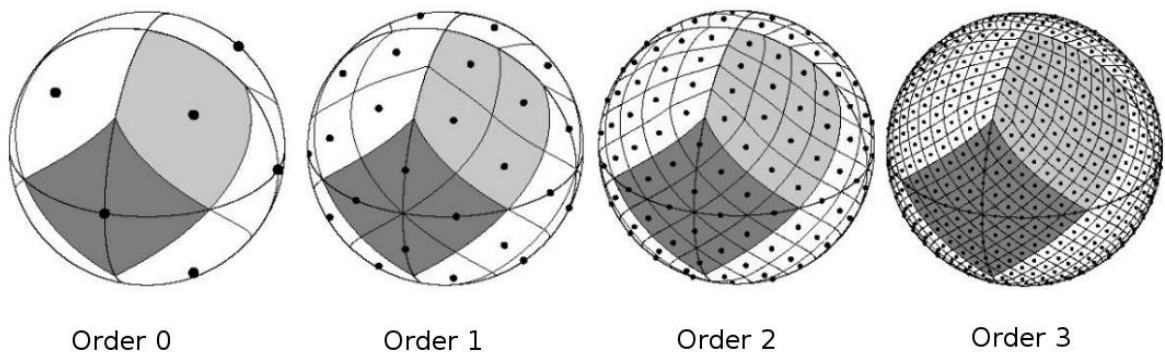

Going beyond its emphasis on visualisation, HiPS can also be described as a generalised spatial mapping of data into a common architecture, and as such HiPS opens the door for a variety of science use cases that involve comparing and combining surveys.

# 3 HiPS principle

HiPS uses the Hierarchical Equal Area isoLatitude Pixelization tessellation technique (HEALPix). The motivations for using the HEALPix tessellation have already been discussed in the MOC IVOA standard 1.0 document in section 1.4 [4]. The main arguments in favour of HEALPix are the equal area property, the fast and constant computation time which does not depend on the hierarchical level, the accuracy and performance of the available libraries (0.4 mas), and the wide usage of HEALPix in several sky missions (WMAP, PLANCK, LIGO…).

HEALPix is a curvilinear partitioning of the sphere that supports a hierarchical tree structure for multi-resolution applications. The detailed geometry and properties of HEALPix are described in Górski et al. 2005 [5]. The HEALPix partitioning of the sphere has a base resolution that divides the sphere into 12 cells, each of which is sub-divided into 4 cells recursively. Thus the sphere at order 1 consists of 48 cells, 192 cells at order 2, 768 at order 3 and so on. Each cell at a given level covers an equal area of the sphere.

| Order 0 | Order 1 | Order 2 | Order 3 |

HiPS is essentially a mapping of survey data at various spatial resolutions into a collection of *HiPS tiles* stored in regular files inside a basic file system.

The *HiPS tiles* define the basic unit of storage. They contain the data (pixels for images, astronomical objects for catalog, 3D-pixels for cubes, etc) spatially organized in the corresponding HEALPix cell. Each *HiPS tile* is fully identified by its HEALPix order and its HEALPix cell index.

The next three sections of this document are normative. Respectively, they present the *HiPS encoding* method (section 4), the *HiPS distribution and registration* protocol (section 5), and the *HiPS client access and use* procedures (section 6). These three aspects of HiPS system and their associated components - *HiPS servers, HiPS clients, registry, HiPS data, HiPS list* - are illustrated in the following picture. These components will be described in the relevant sections.

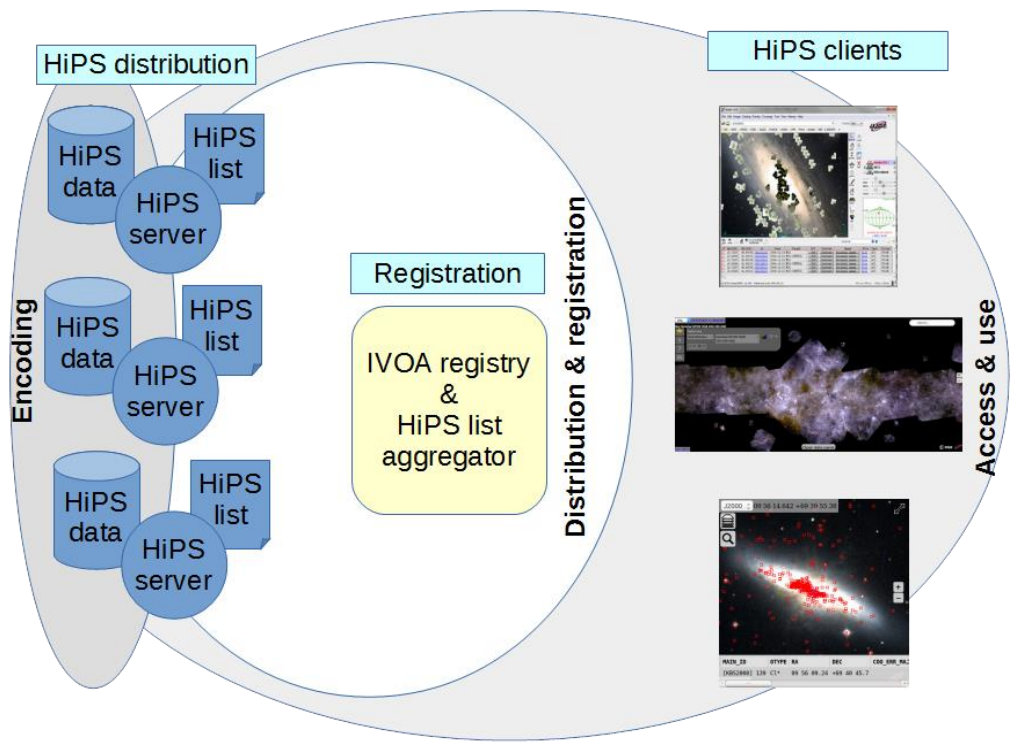

*1 - HiPS components*

In these three sections, the keywords "must", "required", "should", and "may" used in this document are to be interpreted as described in the W3C specifications RFC 2119 [7]. Mandatory elements are indicated as **must**, recommended elements as **should**, and optional elements as **may** or simply "may" without the bold face font.

# 4  HiPS encoding

This section describes the structure of *HiPS data*: the format and the architecture of the *HiPS tiles.* This takes into account the nature of surveys (images, catalogues, cubes, etc) - and the associated *meta data* elements provided with HiPS.

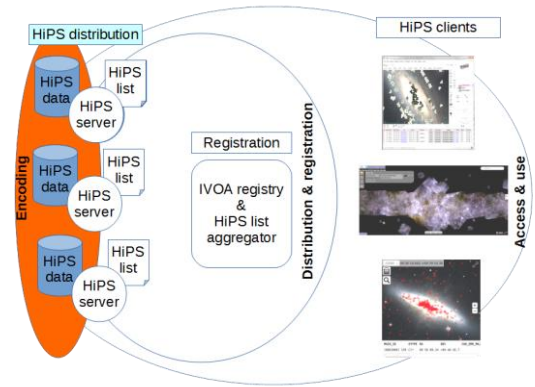

## 4.1  HiPS directory structure

The tiles used by a HiPS **must** be visible as a collection of files in a directory tree. The structure of these directories **must** follow the hierarchy: order → tiles, by using respectively the prefix "Norder" for orders, and "Npix" for tile indices. To avoid directories becoming too large, the tiles **must** be grouped by 10,000 items, using the subdirectory prefix "Dir".

Formally, the tile index *N* in order *K* is located in the HiPS directory structure by this rule:

---

Tile **N** in order **K** → Norder**K**/Dir**D**/Npix**N**{.ext}

  where **D**=(**N**/10000)*10000 (integer division),
  and *{.ext}* is depending of the nature of surveys (see "tile" format below)

---

*Example: the tile containing the HiPS data corresponding to the HEALPix cell index 10302 at the order 6 will be stored in the file Norder6/Dir10000/Npix10302{.ext}*

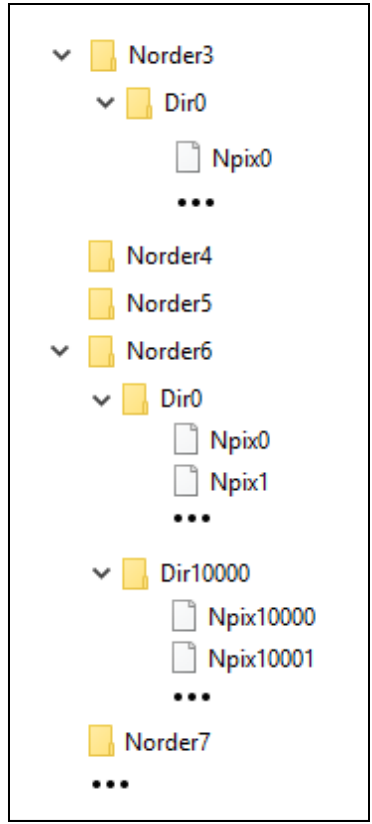

*2 - Example of HiPS directory and file structure*

The spatial location on the sphere of any tile (center of the tile, corners, etc) is directly computed from the tile index using HEALPix formulae [5]. Even if HEALPix supports both "NESTED" and "RING" numbering schemes. HiPS **must** use the NESTED numbering scheme only.

Following the HEALPix geometry, there are four times more tiles at order K+1 compared to the order K.The tile index N at order K corresponds to the 4 tile indices Nx4, Nx4+1, Nx4+2 and Nx4+3 at order K+1.

The actual implementation of HiPS as directories and files is not an obligation, only the view as directories and files is required (see HiPS distribution section). Internally, a HiPS **may** be stored in a data base, or any other appropriate packaging (tar or zip files…) rather than in a basic file system directory structure.

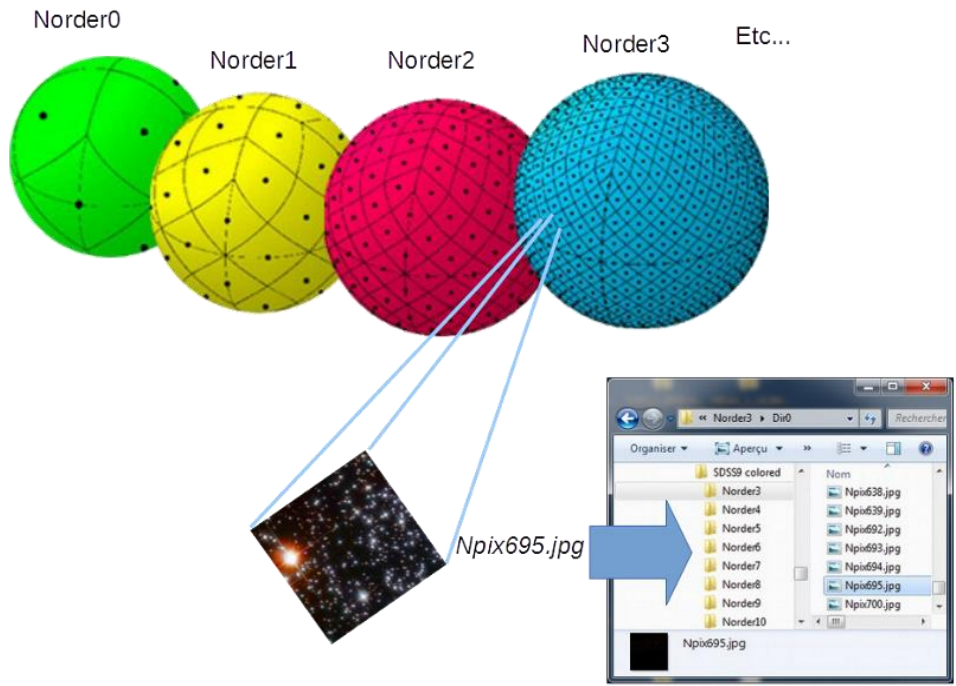

*3 - HiPS architecture based on directories and files*

The number of HiPS orders depends on the original data parameters: the angular resolution for a pixel survey, the number of sources for a catalogue, etc. It also depends on the size of the *HiPS tile* (see next section). A HiPS **may** use a partial tree (where not all sphere HEALPix cells are described), notably for a survey that does not cover the whole sky, and it **may** be possibly not uniform (different depth for different branches) like for catalogue with different densities on the sphere.

## *4.2 HiPS tile formats*

The content of the *HiPS tiles* depends on the nature of the original survey: pixel arrays for an image survey, catalogue source list for a catalogue survey, cube arrays for a cube survey, vector arrays for polarization data, properties for localized meta-data (progenitors), etc. This document describes only the format of the tiles dedicated to images, catalogues and cubes corresponding respectively to *image HiPS*, *catalogue HiPS*, and *cube HiPS*.

A *HiPS tile* **must** contain the data (pixels, catalogue sources...) located in its associated HEALPix cell on the sky. This implies that the data **must** have a footprint on celestial sphere (described by a WCS solution for images or cubes, and described by spherical coordinates for catalogue sources). The tile format depends on the survey data type: FITS, JPEG, PNG for image or cube surveys; TSV for catalogues. These basic tile formats have been especially chosen in order to facilitate checking of the tile content with basic file tools and editors.

### 4.2.1  Image HiPS tile

 *Image HiPS tiles* **must** contain the image pixels located in the associated HEALPix cell. This constraint requires a resampling of the original image pixels into the HEALPix geometry grid (see next section), except for cases where the original data are natively provided in HEALPix (for instance PLANCK data),

To avoid tiles containing only one HEALPix pixel, HiPS image tile hierarchy is S orders less deep than the original HEALPix resampled data, packaging the $2^S$ x $2^S$ HEALPix cell values, as an array of pixels.

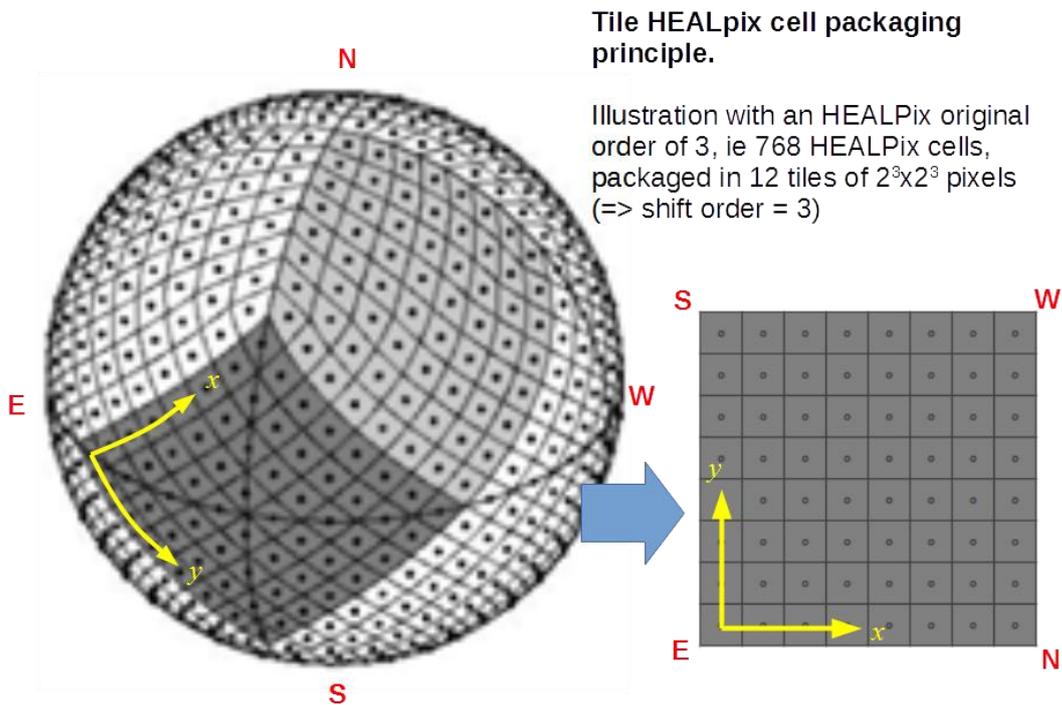

**Tile HEALpix cell packaging principle.**

Illustration with an HEALPix original order of 3, ie 768 HEALPix cells, packaged in 12 tiles of $2^3$x$2^3$ pixels (=> shift order = 3)

*4 - The HEALPix cell packaging principle (in a FITS tile – see 4.2.2.1)*

*Example: The XMM PN survey (4"pixel angular resolution), requiring a resampling in a HEALPix grid at order 16 (3.221" HEALPix cell angular resolution), may be packaged in HiPS tile at order 7 (16-9) containing the $2^9$ x $2^9$ HEALPix cell values.*

<u>In practice</u>: *because of present capacities of most network infrastructure a tile size of 512x512 pixels (shift order: S = 9) is a good compromise between the size of the tiles and the number of tiles required to map the original survey data.*

### 4.2.1.1  Deepest HiPS order

At the deepest HiPS order, the pixels in the tiles are directly computed from the original data. The method for evaluating each pixel value of these tile arrays depends on the HiPS creation algorithm. It **may** be the nearest pixel, a bilinear interpolation or any other algorithm mapping the original pixels from their original sky grid into the HEALPix grid.

Also, in the case of overlapping multiple input original images, an average method, weighted or not, can be applied. This document does not recommend the usage of any specific algorithm.

If the original data come from a native HEALPix map (for instance, PLANCK data), no pixel conversion is then required: each HiPS tile pixel takes the associated HEALPix map cell value.

The choice of the deepest HiPS order depends on the required HiPS resolution. It **may** be chosen as the first HEALPix order under the original pixel angular resolution. This choice depends on the tile width according to this HEALPix function:

*Tile pixel angular size* =~ sqrt( 4*PI / (12 x (*tileWidth* x $2^{order}$)$^2$) )

| Deepest HiPS Order | Number of tiles | Tile angular size | Tile pixel angular size |
|---|---|---|---|
| 0 | 12 | 58.63° | 6.871' |
| 1 | 48 | 29.32° | 3.435' |
| 2 | 192 | 14.66° | 1.718' |
| 3 | 768 | 7.329° | 51.53″ |
| 4 | 3072 | 3.665° | 25.77″ |
| 5 | 12288 | 1.832° | 12.88″ |
| 6 | 49152 | 54.97' | 6.442″ |
| 7 | 196608 | 27.48' | 3.221″ |
| 8 | 786432 | 13.74' | 1.61″ |
| 9 | 3145728 | 6.871' | 805.2mas |
| 10 | 12582912 | 3.435' | 402.6mas |
| 11 | 50331648 | 1.718' | 201.3mas |
| 12 | 201326592 | 51.53″ | 100.6mas |
| 13 | 805306368 | 25.77″ | 50.32mas |
| 14 | 3221225472 | 12.88″ | 25.16mas |
| 15 | 12884901888 | 6.442″ | 12.58mas |
| 16 | 51539607552 | 3.221″ | 6.291mas |
| 17 | 2,06158E+11 | 1.61″ | 3.145mas |
| 18 | 8,24634E+11 | 805.2mas | 1.573mas |
| 19 | 3,29853E+12 | 402.6mas | 786.3µas |
| 20 | 1,31941E+13 | 201.3mas | 393.2µas |

*5 - Tile pixel angular size function of the HiPS order with 512x512 tile size*

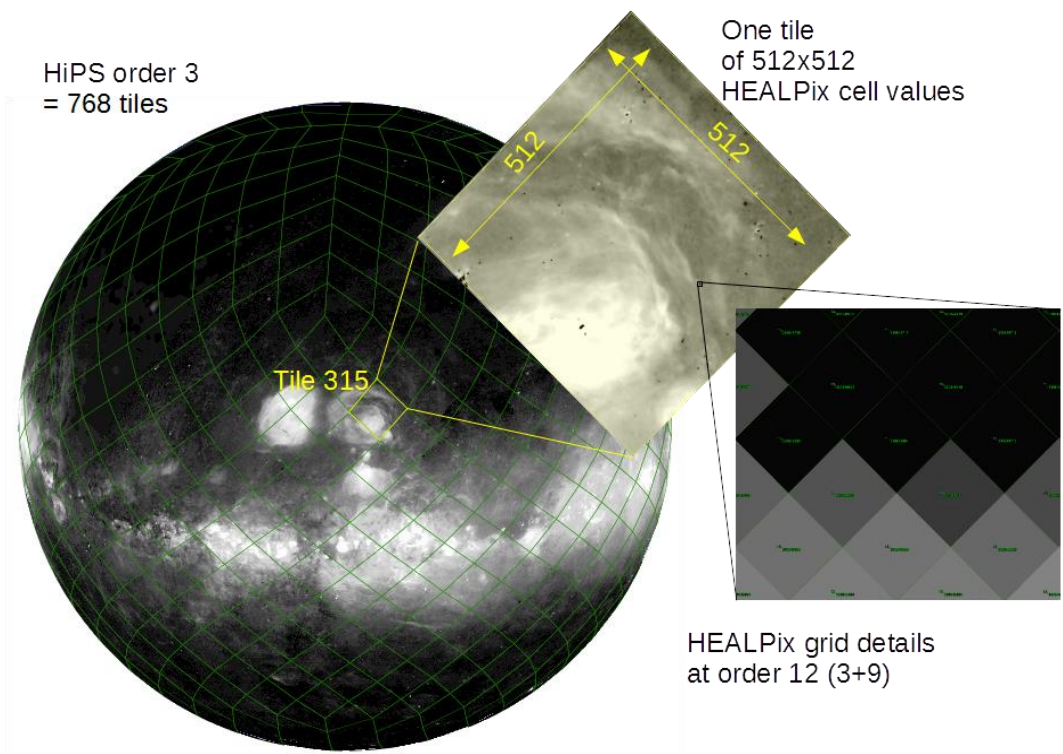

6 - *The HiPS Halpha Finkbeiner survey at HiPS order 3 (768 tiles) with HiPS tiles of 512x512 HEALPix cell values (51.53" resolution)*

### 4.2.1.2  Other HiPS orders

As explained in the tile structure, there are four times less tiles at order K compared to the order K+1. Each pixel of the tiles at order K is computed from the four pixels at order K+1 covering the same HEALPix area. The method for evaluating this pixel from its 4 sub-pixels depends on the HiPS creation algorithm. It **may** be the first pixel (one value amongst the 4 values), the mean, the median value, etc. This document does not recommend the usage of any specific algorithm.

*In practice: the mean method is appropriate when the original pixel values are representing some intensity by unit of surface, the median method highlights large scale structures and reduce the impact of bright stars, and the first pixel is the fastest method.*

### 4.2.1.3  Format of tiles

Three image formats **may** be used to package the *HiPS tiles for images*: FITS [6], PNG or JPEG. An image tile **may** be stored in FITS format in order to keep the full dynamic range of the original data values. Tiles stored in JPEG format provide good file compression, and tiles stored in PNG format provides the capability to support transparency channel. The tile file extension **must** correspond to the format: *.fits* for FITS, *.jpg* for JPEG, *.png* for PNG. These extensions **must** be in lowercase.

An image HiPS **may** be delivered simultaneously in various formats.

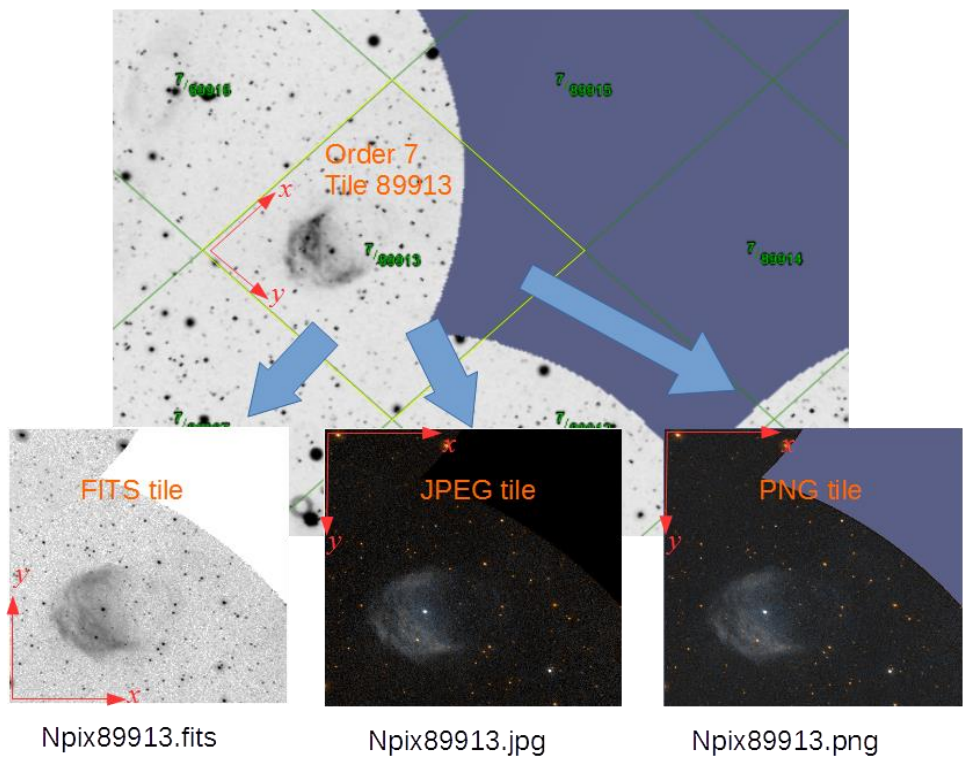

Npix89913.fits  Npix89913.jpg  Npix89913.png

*7 – A GALEX HiPS survey tile provided simultaneously in FITS, JPEG and PNG*

<u>Note</u>: Contrary to the FITS convention, in JPEG and PNG the lines of the pixel array are stored in top->down direction.

### 4.2.2 Catalogue HiPS tile

A *catalogue HiPS tile* **must** contain a list of catalogue sources corresponding to the original sources located in the associated HEALPix cell.

The distribution of the sources in the HiPS hierarchy **may** use various algorithms. The most usual method is based on the following principle: if the number of sources in a tile at order K overtakes a threshold, the other sources are stored in the 4 sub-tiles of the order K+1, and so on, recursively. This threshold is not necessary a constant, it **may** depend of the distribution of the sources, notably for handling the possible local densities. The method for sorting the source **may** be based on any property of the sources: magnitude, parallax, number of citations…

<u>*In practice:*</u> *a HiPS tile containing a few hundred sources is a good compromise between the size of the tile, the time used to display the information, and the number of tiles required to map the catalogue.*

The deepest HiPS order for a catalogue HiPS is related to the number and the density of the sources and the number of sources stored in each tile.

#### 4.2.2.1 Format of tiles

A *catalogue HiPS tile* **must** be stored in an UTF-8 file (ie ASCII 7 bits + extended characters [11]) using this TSV convention (Tab Separated Value). The first line is a header providing the column names. All fields, are separated by a TAB (decimal ASCII code 9), and each line is ended by a LF (decimal ASCII code 10), possibly preceded by a CR (decimal ASCII code 13). Any line beginning with a hash (# - decimal ASCII code 35) is a comment line and it is ignored. An empty field corresponds to the null value (not provided).

All tiles **must** have the same columns and the same column ordering. Columns providing the location on the sky of each source **must** be present (see 4.4.3). The tile file extension **must** be *".tsv"* in lowercase.

```
# Progressive catalog: I/317/sample; version date: 2016-06-06T07:49Z
PPMXL_ID→RAJ2000→DEJ2000→errHalfMaj→errHalfMin→errPosAng→Jmag→Hmag→Kmag→e_Jm
850851035651566199→190.539132→+48.805653→0.015→0.015→90→10.985→10.807→10.806
856098854447350345→192.177591→+47.775121→0.063→0.063→90→14.111→13.567→13.238
850782239565659836→191.315926→+48.718502→0.015→0.015→90→11.013→10.769→10.699
857096068483317360→189.754712→+48.399660→0.019→0.015→90→11.093→10.794→10.780
850899123912864590→190.428803→+49.341103→0.047→0.045→90→10.513→9.961→9.872→0.
856197121630867029→192.267307→+47.528337→0.015→0.015→90→11.315→11.117→11.099
850915003259720486→190.113383→+49.140826→0.036→0.031→90→10.706→10.453→10.351
856615640088063651→192.130212→+46.965377→0.017→0.015→90→11.172→10.934→10.859
850879950281314873→190.788816→+48.325225→0.018→0.015→0→11.154→10.835→10.786→
856896865603869199→190.659080→+48.103347→0.016→0.015→0→11.003→10.811→10.733→
856815965863606702→191.661529→+48.186958→0.019→0.018→90→10.986→10.674→10.614
850876367388720869→190.978421→+48.312087→0.015→0.015→90→10.977→10.688→10.609
850838827165173147→191.035848→+48.644478→0.063→0.063→90→13.507→12.732→12.324
```

*8 - Excerpt of the catalog HiPS tile Norder5, Npix2741 from PPMXL*

### 4.2.3 Cube tile format

Cube HiPS tiles are based on the same format as image HiPS tiles. Cube tiles are sliced in multiple frames saved in image HIPS tiles. The number of frame tiles **may** correspond to the depth of the original cubes.

Each of the frame tile name **must** have a suffix "_n" where "n" is the depth index within the cube, i.e. the frame number. The tile name without the suffix corresponds to the first frame (depth = 0). This way a HiPS client which is not cube HiPS compatible will still be able to display at least the first frame of the cube HiPS as an image HiPS. As with image HiPS, the format of each cube frame tiles **may** be FITS, JPEG and/or PNG.

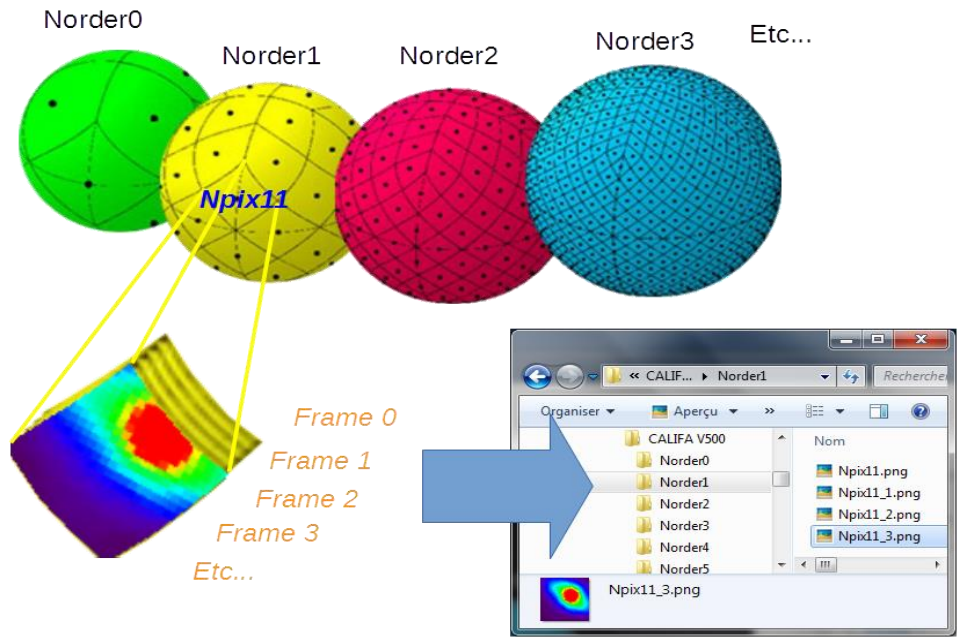

*9 - Principle of the cube HiPS tiles*

## 4.3 Low resolution "enablers"

The drawing of the lowest HiPS orders that correspond to large sky regions (>30 deg) can be difficult because of:

    a) the high level of distortion of these tiles when a basic/fast 4 corners bilinear drawing method is used (see section 6).

    b) the large number of tiles required for drawing large regions, even at low HiPS order.

To ease the client drawing process, two enablers may be implemented for the lower orders: order omission and Allsky preview.

### 4.3.1 Order omission

The low orders: order 0 (12 tiles), order 1 (48 tiles), order 2 (192 tiles) **may** be simply omitted, meaning that the survey files are not provided at these low resolutions.

### 4.3.2 *Allsky* preview file

The tiles at low orders (0 to 3) may be packaged together into a unique file called *Allsky*. These files **must** be located in the *NorderK* corresponding directory (where K is between 0 and 3). The associated regular tiles **must not** be removed, notably for supporting basic HiPS clients. The method to generate the *Allsky* file depends on the nature of surveys:

    • *Image HiPS Allsky:* The *Allsky* file is built as an array of tiles, stored side by side in the left-to-right order. The width of this array **must** be the square root of the number of the tiles of the order. For instance, the width of this array at order 3 is 27 ( (int)sqrt(768) ). To avoid having a too large *Allsky* file, the resolution of each tile **may** be

reduced but **must** stay a power of two (typically 64x64 pixels rather than 512x512).

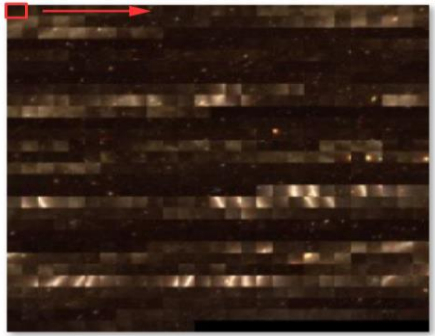

*10 - Example of HiPS Fermi color Allsky.jpg file*

- *Catalogue HiPS Allsky:* The *Allsky* file is built as the concatenation of all the catalogue tiles.

- *Cube HiPS Allsky:* The *Allsky* file associated to cube HiPS follows the same rule than the regular cube frame tiles one: it is divided in *Allsky_n* files for providing each frame separately. Without suffix, the *Allsky* file corresponds to the frame 0 and each individual *Allsky_n* files do have to follow the same rules as the Image HIPS *Allsky* specified above.

## *4.4  Meta data*

Five additional files **must**, **should** or **may** be used for specifying survey meta data:

1. the *properties* file (required);
2. the *Moc.fits* file (recommended under condition);
3. the *metadata* file (recommended under condition);
4. the *preview.jpg* file (optional);
5. the *index.html* file (optional).

### 4.4.1  The *properties* file

A text file named *properties* **must** be provided. It specifies generic meta information such as identification, description, copyright, creation date, pixel range, etc.

This *properties* file **must** be located in the HiPS root directory. It **must** be coded in UTF-8 (ie ASCII 7 bits + extended characters [11]), one line per property, following the syntax *keyword = value*. Each line is ended by LF (decimal ASCII code 10), possibly preceded by CR (decimal ASCII code 13). Lines starting with a hash as first character (decimal ASCII code 35) **must** be ignored as comment line. Blank lines **must** be ignored. Key word cannot include the character "=" (decimal ASCII code 61) or spaces (blank, TAB, ..), and the possible spaces before and after the character "=" **must** be ignored. The ordering of the keywords is not important. The number of characters in values is not limited.

```
creator_did           = ivo://SSC/P/XMM/EPIC
obs_collection        = XMM-Newton
obs_title             = XMM-Newton stacked EPIC images
obs_ack               = HE-Team (Strasbourg), SSC XMM-Newton
obs_copyright         = (c) ESA / SSC XMM-Newton
obs_copyright_url     = http://xmmssc.irap.omp.eu/
t_min                 = 51577.46
t_max                 = 56331.92
obs_regime            = X-ray
em_min                = 1e-10
em_max                = 6e-9
s_pixel_scale         = 0.001111
data_pixel_bitpix     = 32
dataproduct_type      = image
moc_sky_fraction      = 0.1404
hips_estsize          = 40322637
hips_creation_date    = 2015-10-01T14:07Z
client_category       = Image/X/XMM
hips_builder          = Aladin/HipsGen v8.175
hips_version          = 1.4
hips_release_date     = 2015-10-05T08:46Z
hips_publisher        = SSC XMM-Newton
hips_frame            = equatorial
hips_order            = 7
hips_tile_width       = 512
hips_master_url       = http://saada.u-strasbg.fr/xmmallsky
hips_status           = public master clonable
hips_tile_format      = png fits
hips_pixel_bitpix     = 32
hips_hierarchy        = mean
hips_pixel_cut        = 0 50.95
hips_data_range       = -266.5 799.5
hips_pixel_scale      = 8.946E-4
```

*11 - Example of a properties file*

<u>Note:</u> The keyword vocabulary re-uses as far as possible the ObsCore IVOA vocabulary and syntax [8].

There are 9 mandatory keywords which **must** be specified:

1. **creator_did**: Unique identifier of the HiPS – Format: IVOID [3] – e.g.: ivo://CDS/P/2MASS/J

2. **obs_title**: Data set title – Format: free text, but should be short (no more than one line) – e.g.: HST 110W

3. **dataproduct_type**: Type of data – Format: word "*image*" "*catalog*", "*cube*" respectively for image HiPS, catalogue HiPS, cube HiPS.

4. **hips_version**: HiPS version number – Format: word "*1.4*" corresponds to this document specification.

5. **hips_release_date**: HiPS release date – Format: ISO 8601 (YYYY-mm-ddTHH:MMZ)

6. **hips_status**: HiPS status description – Format: list of blank separated words ("private" or "public"), ("master", "mirror", or "partial"), ("clonable",

"unclonable" or "clonableOnce"[1]) – Default : public master clonableOnce

7. **hips_tile_format**: Tile formats – Format: list of different HIPS tiles format supported by the survey, space separated (one or many of "fits", "jpeg", "png" for image/cube HiPS and "tsv" for catalog HiPS)
8. **hips_order**: Deepest HiPS order – Format: positive integer
9. **hips_frame**: Coordinate frame reference – Format: "*equatorial*" (ICRS), "*galactic*", "*ecliptic*" according to the used HiPS coordinate reference.

There are 2 additional mandatory keywords specific to the type of HiPS, and they **must** be specified in the following cases:
1. **dataproduct_subtype**: Subtype of data – Format: word "*color*", "*live*" respectively for image HiPS colored (based on colored tiles), and live HiPS (HiPS for which the content may evolve – for instance live databases like Simbad, or observation missions still in progress)
2. **hips_cube_depth**: Number of frames of the HiPS cube – Format: positive integer

Other keywords **should** or **may** be specified. The list below is not exhaustive and new keywords **may** be added if required. The vocabulary associated to some keywords are not exhaustive and **may** be extended if required. Some keywords **may** be repeated to specify multiple values like provenance (suggested by (*) in the following table)

| Keyword | Req or Should | Description – Format - Example |
|---|---|---|
| creator_did | R | Unique ID of the HiPS - Format: IVOID - Ex : ivo://CDS/P/2MASS/J |
| publisher_id | | Unique ID of the HiPS publisher – Format: IVOID - Ex : ivo://CDS |
| obs_collection | | Short name of original data set – Format: one word – Ex : 2MASS |
| obs_title | R | Data set title – Format: free text, one line – Ex : HST F110W observations |
| obs_description | S | Data set description – Format: free text, longer free text description of the dataset |
| obs_ack | | Acknowledgment mention. |
| prov_progenitor (*) | S | Provenance of the original data – Format: free text |
| bib_reference (*) | | Bibliographic reference |
| bib_reference_url (*) | | URL to bibliographic reference |
| obs_copyright | | Copyright mention associated to the original data – Format: free text |
| obs_copyright_url | | URL to a copyright mention |
| hips_copyright | | Copyright mention associated to the HiPS – Format: free text |
| obs_regime (*) | S | General wavelength – Format: word: "Radio" \| "Millimeter" \| "Infrared" \| "Optical" \| "UV" \| "EUV" \| "X-ray" \| "Gamma-ray" |
| data_ucd (*) | | UCD describing data contents |
| hips_version | R | Number of HiPS version – Format: 1.4 (corresponds to this document) |

---

[1] "clonableOnce" implies that the copy is allowed, but only from the "master" site

| | | |
|---|---|---|
| **hips_builder** | | Name and version of the tool used for building the HiPS – Format: free text |
| **hips_creator** | | Institute or person who built the HiPS – Format: free text – Ex : CDS (T.Boch) |
| **hips_creation_date** | S | HiPS first creation date - Format: ISO 8601 => YYYY-mm-ddTHH:MMZ |
| **hips_release_date** | R | Last HiPS update date - Format: ISO 8601 => YYYY-mm-ddTHH:MMZ |
| **hips_service_url** | | HiPS access url – Format: URL |
| **hips_status** | R | HiPS status – Format: list of blank separated words (private" or "public"), ("master", "mirror", or "partial"), ("clonable", "unclonable" or "clonableOnce") – Default : public master clonableOnce |
| **hips_estsize** | | HiPS size estimation – Format: positive integer – Unit : KB |
| **hips_frame** | R | Coordinate frame reference – Format: word "equatorial" (ICRS), "galactic", "ecliptic" |
| **hips_order** | R | Deepest HiPS order – Format: positive integer |
| **hips_tile_width** | | Tiles width in pixels – Format: positive integer – Default : 512 |
| **hips_tile_format** | R | List of available tile formats. The first one is the default suggested to the client – Format: list of word blank separated: "jpeg", "png", "fits", "tsv" |
| **hips_pixel_cut** | | Suggested pixel display cut range (physical values) – Format: min max – Ex : 10 300 |
| **hips_data_range** | | Pixel data range taken into account during the HiPS generation (physical values) – Format: min max – Ex : -18.5 510.5 |
| **hips_sampling** | | Sampling applied for the HiPS generation – Format: words "none", "nearest", "bilinear" |
| **hips_overlay** | | Pixel composition method applied on the image overlay region during HiPS generation – Format: word "add", "mean", "first", "border_fading", "custom" |
| **hips_skyval** | | Sky background subtraction method applied   during HiPS generation – Format: word: "none", "hips_estimation", "fits_keyword" |
| **hips_pixel_bitpix** | | Fits tile BITPIX code – Format: -64, -32, 8, 16, 32, 64 (FITS convention) |
| **data_pixel_bitpix** | | Original data BITPIX code   - Format: -64, -32, 8, 16, 32, 64 (FITS convention) |
| **dataproduct_type** | R | Type of data – Format: word "image", "cube", "catalog" |
| **dataproduct_subtype** | R | Subtype of data – Format: word "color", "live" |
| **hips_progenitor_url** | | URL to an associated progenitor HiPS |
| **hips_cat_nrows** | S | Number of rows of the HiPS catalog – Format: positive integer |
| **hips_cube_depth** | R | Number of frames of the HiPS cube – Format: positive integer |
| **hips_cube_firstframe** | | Initial first index frame to display for a HiPS cube – Format: positive integer – Default : 0 |
| **data_cube_crpix3** | | Coef for computing physical channel value (see FITS doc) – Format: real |
| **data_cube_crval3** | | Coef for computing physical channel value (see FITS doc) – Format: real |
| **data_cube_cdelt3** | | Coef for computing physical channel value (see FITS doc) – Format: real |
| **data_cube_bunit3** | | Third axis unit (see FITS doc) – Format: string |
| **hips_initial_ra** | S | Default RA display position – Format: real (ICRS frame) – Unit : degrees |
| **hips_initial_dec** | S | Default DEC display position – Format: real (ICRS frame) – Unit : degrees |
| **hips_initial_fov** | S | Default display size – Format: real – Unit : degrees |
| **hips_pixel_scale** | | HiPS pixel angular resolution at the highest order – Format: real – Unit : degrees |

| s_pixel_scale | | Best pixel angular resolution of the original images – Format: real – Unit : degrees |
|---|---|---|
| t_min | S | Start time of the observations – Format: real – Representation: MJD[2] |
| t_max | S | Stop time of the observations – Format: real – Representation: MJD |
| em_min | S | Start in spectral coordinates – Format: real – Unit: meters |
| em_max | S | Stop in spectral coordinates – Format: real – Unit: meters |
| client_category | | '/' separated keywords suggesting a display hierarchy to the client – Ex : Image/InfraRed |
| client_sort_key | | Sort key suggesting a display order to the client inside a "client_category" – Format: free text – Sort : alphanumeric |
| addendum_did(*) | | In case of "live" HiPS, creator_did of the added HiPS |
| moc_sky_fraction | | Fraction of the sky covers by the MOC associated to the HiPS – Format: real between 0 and 1 |

### 4.4.2 The *Moc.fits* file

A file named *Moc.fits*, located in the HiPS root directory, **should** be provided if the HiPS coordinate reference is equatorial, or if the data covers the whole sky, and **may** be provided otherwise. It represents the coverage map of the survey (see IVOA MOC REC 1.0). This file follows the IVOA MOC standard [4]. The client can use it as a summary of the coverage of the HiPS to avoid trying to load tiles outside the coverage area of the HiPS.

Note: MOC is also based on HEALPix, and it only supports equatorial coordinate reference system. Thus, the MOC generation from an equatorial HiPS is straightforward due to the fact that the MOC can be directly built from the list of HiPS tile indices. If the HiPS is based on another coordinate system the way to generate the HiPS may be complex. The special case of the whole sky is simply a MOC that covers the whole sky (no longer coordinate system dependent).

### 4.4.3 The *metadata* file

In addition to the generic meta data described in the *properties* file (see 4.4.1), a file called *metadata.xxx*, located in the HiPS root directory **may** be provided for specific meta information such as column descriptions, original FITS keywords, etc. This information depends on the survey data type:

- *Image HiPS metadata:* the meta data are stored as a FITS header (FITS convention) in *metadata.fits* file for providing generic FITS header keywords. This file **can** be reduced to a FITS header or even a regular FITS file from the original survey containing the FITS header used for global metadata information – in the second case the image array is simply ignored by the client.

- *Catalogue HiPS metadata:* the meta data are stored in *metadata.xml* file as a fully defined VOTable (column descriptions, units, ucds,...) [9]. All column information associated with the source tiles **must** be provided via the "metadata" file. The internal header of each individual tile is ignored by the client (VOTable meta data richer than the default

---

2    . MJD date can be computed from Unix time: MJD = (Unix_time / 86400) + 40587

simple header line, notably for designating columns containing the spatial locations).

- *Cube HiPS metadata:* same as for HiPS image.

### 4.4.4 The *preview.jpg* file

A file called *preview.jpg*, located in the HiPS root directory **may** be provided to offer a preview of the HiPS. This image is 256 x 256 pixel size, stored in JPEG format.

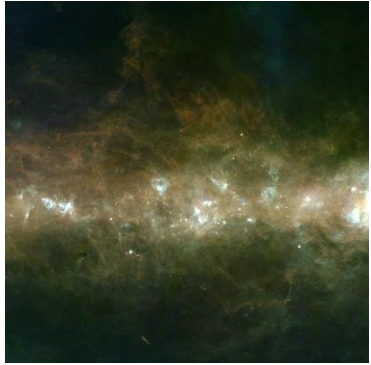

*6 - Example of preview.jpg file associated to the AKARI colored HiPS*

### 4.4.5 The *index.html* file

A file called *index.html*, located in the HiPS root directory **may** be provided to offer a simple HTML presentation of the survey. By loading the root HiPS directory location one will be able to display information about the HiPS itself through any Web browser.

*7 - Example of index.html file generated by Aladin/HipsGen tool displayed in a Web browser*

# 5 HiPS distribution and registration

This section describes the notions of *HiPS server*, *HiPS list* and HiPS registration required for the building of the HiPS distribution system.

A HiPS may be used locally by a dedicated client, in which case no distribution or registration mechanism are required. However HiPS are usually published on-line so that remote clients can access them. These clients need to know the existence and the location of the available HiPS. It is also possible to have duplicated HiPS at different locations to ensure continuity and faster access.

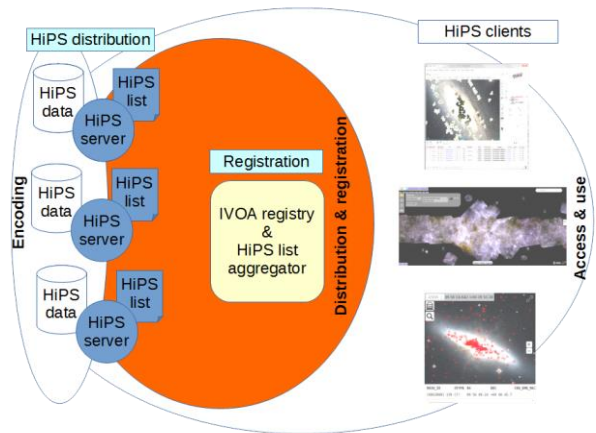

## 5.1 HiPS server

A *HiPS server* is a regular HTTP site which provides HiPS surveys. As HiPS are seen as a simple hierarchy of directories and files, the distribution of HiPS **may** be operated by publishing the HiPS file structures via the HTTP server. However the actual implementation of HiPS as directories and files is not an obligation, only the view as directories and files is required. Behind the HTTP server, a HiPS **may** be stored in a database, or any other appropriate method for packaging it (tar or zip files…) rather than a basic file system directory.

Note: According to the HTTP site configuration, the tiles, notably the FITS tiles, **may** or **may not** be compressed. So it is up the clients to ensure that the required uncompression step is performed (as can be done transparently by the HTTP libraries)

Additionally, a *HiPS server* **must** implement a dedicated URL which provides the list of the HiPS surveys following the *HiPS list* syntax (see below).

## 5.2 HiPS list

A *HiPS list* is the list of HiPS surveys published by a given HiPS server. The *HiPS list* **must** be provided by any *HiPS server* via a dedicated URL.

The format of this list is derived from the HiPS *properties* file associated to each HiPS. It is an UTF-8 file containing a collection of records (key1 = property1\nkey1 = property2...) that are separated by blank lines and have the same syntax and vocabulary as the HiPS *properties* file.

Each record of the HiPS list **must** provide at least these four mandatory properties: *creator_did, hips_release_date, hips_service_url, hips_status*. Other HiPS properties **may** or **may not** be provided. See the properties table in 4.4.1.

```
# Hipslist of http://alasky.u-strasbg.fr HiPS server
creator_did        = ivo://CADC/P/HST/F850LP/r3
hips_release_date  = 2014-10-14T12:00Z
hips_service_url   = http://alasky.u-strasbg.fr/HST/F850LP
hips_status        = public master clonable

creator_did        = ivo://CDS/P/2MASS/H
hips_release_date  = 2014-11-03T12:00Z
hips_service_url   = http://alasky.u-strasbg.fr/2MASS/H
hips_status        = public mirror unclonable
hips_estsize       = 1610612736
hips_order         = 9
hips_tile_format   = fits jpeg
dataproduct_type   = image
moc_sky_fraction   = 1
```

*14 - Example of a HiPS list*

<u>In practice:</u> a HiPS list **may** be easily generated as the concatenation of the properties files[3] of all distributed HiPS, with blank line separators.

## 5.3 HiPS registration

HiPS surveys and/or servers **may** be registered in the IVOA registry [2].

HiPS registration in the IVOA registry **may** occurs at two different levels:

1. *Individual HiPS survey* registration could be useful to at least get a valid IVOID identifier [3]. However this step in not required for client visibility;

2. *HiPS server* registration allowing HiPS clients to discover all the HiPS via their HiPS list.

Only the *HiPS server* registration is required in order to be visible through compatible clients. The HiPS clients get the HiPS list for each registered HiPS server in order to compute the list of available HiPS surveys, their location, and their potential mirror sites.

<u>In practice:</u> this task may be done by an intermediate registry tool (HiPS list aggregator) which maintains an up-to-date merging of all available HiPS lists[4]. This approach is similar to the Global TAP schema aggregator described in section 3.1 of the IVOA "Discovering Data Collections" note [10].

---

[3] In this case, the possible blank lines inside a properties file must be removed.

[4] An example of this aggregator service can be used thanks to these URLs: ASCII result: http://aladin.u-strasbg.fr/hips/globalhipslist, JSON result: http://aladin.u-strasbg.fr/hips/globalhipslist?fmt=json

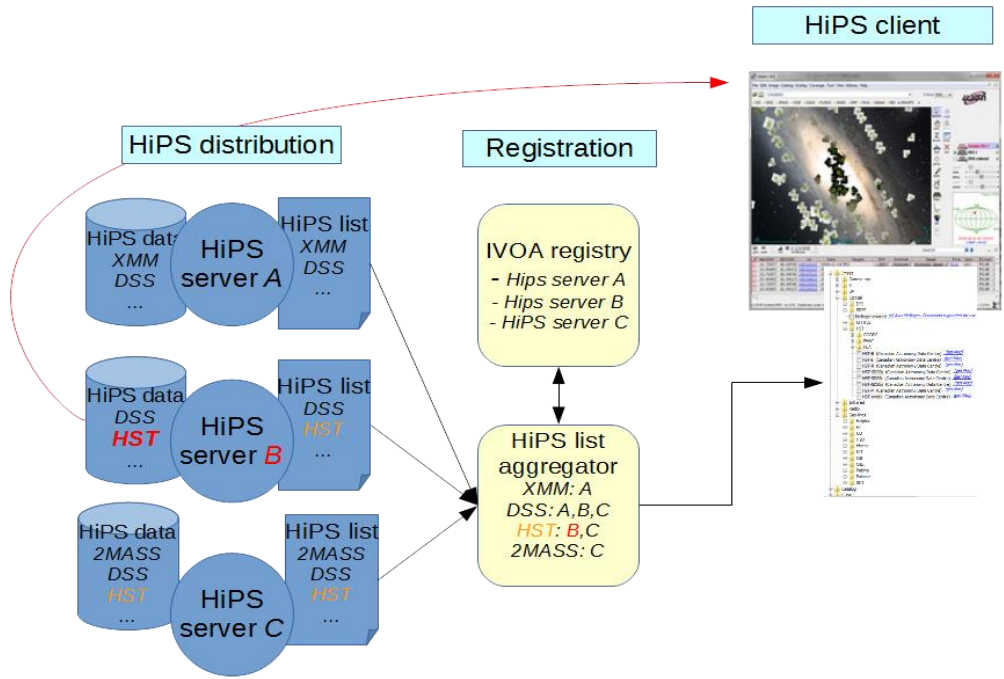

*8 - HiPS client discovering principle*

## 5.3.1 HiPS server registration

A HiPS server **may** be registered in the IVOA registry by using an IVOA XML record declaration. This record **must** provide the capability of the *HiPS list* URL associated to the declared *HiPS server* by using the *standardID*: "*ivo://ivoa.net/std/hips#hipslist-1.0*"

```
<ri:Resource xmlns:(...) xsi:type="vr:Service">
    <title>CDS HiPS Service (master server)</title>
    <shortName>CDS hipsmaster</shortName>
    <identifier>ivo://CDS/hipsmaster</identifier>
    <curation>
        <publisher>CDS</publisher>
        <creator>
            <name>Centre de Donnees astronomiques de strasbourg</name>
        </creator>
        <contact>
            <name>CDS Helpdesk</name>
            <email>cds-question@unistra.fr</email>
        </contact>
    </curation>
    <content>
        <subject>CDS Hierarchical Progressive Surveys</subject>
        <description>The CDS provides a collection of reference
          surveys available thanks to HiPS protocol...</description>
        <referenceURL>http://aladin.u-strasbg.fr/hips</referenceURL>
        <type>Archive</type>
        <contentLevel>Research</contentLevel>
    </content>
    <capability standardID="ivo://ivoa.net/std/hips#hipslist-1.0">
        <interface role="std" xsi:type="vs:ParamHTTP">
          <accessURL use="full" >
            http://alasky.u-strasbg.fr/hipslist</accessURL>
        </interface>
    </capability>
</ri:Resource>
```

*16 - Example of a HiPS server registry record*

### 5.3.2 Individual HiPS survey declarations

Optionally, each HiPS **may** be registered individually in the IVOA registry. There are two constraints: 1 – only the main HIPS creator is authorized to register them (and not by the potential clone site manager) 2 - Only the URL of the original HiPS (master site, not the clone site) **may** be provided by using the *standardID*: *ivo://ivoa.net/std/hips#hips-1.0*

```
 <ri:Resource xmlns:(...) xsi:type="vs:CatalogService">
   <title>ALADIN image DSS2 blue survey collection</title>
   <shortName>Aladin DSS2 blue</shortName>
   <identifier>ivo://CDS/P/DSS2/blue</identifier>
   <curation>
     <publisher>CDS</publisher>
     <creator ivo-id="ivo://CDS">
       <name>Centre de Donnees astronomiques de strasbourg</name>
     </creator>
     <contact>
        <name>CDS Helpdesk</name>
        <email>cds-question@unistra.fr</email>
     </contact>
   </curation>
   <content>
     <subject>ALADIN image DSS2 blue survey collection</subject>
     <description> ALADIN image server provides reference ... </description>
     <referenceURL>http://aladin.u-strasbg.fr</referenceURL>
     <type>Survey</type>
     <contentLevel>Research</contentLevel>
   </content>
   <capability standardID="ivo://ivoa.net/std/hips#hips-1.0">
     <interface role="std" xsi:type="vs:ParamHTTP">
       <accessURL use="base">
        http://alasky.u-strasbg.fr/DSS/DSS2-blue-XJ-S</accessURL>
     </interface>
   </capability>
   <coverage>
      <footprint ivo-id="ivo://mocivod">
         http://alasky.u-strasbg.fr/DSS/DSS2-blue-XJ-S/Moc.fits
      </footprint>
   </coverage>
 </ri:Resource>
```

*17 - Example of an individual HiPS survey registry record*

## 5.4 HiPS mirroring

HiPS surveys are governed by usual data rights policies. The data may be public or private. It may be mirrored or not, according to the original data rights rules set by the HiPS creator.

If both the original data copyright statement and the HiPS creator authorize the duplication and the distribution of any derived HiPS products, it could be interesting to mirror HiPS in order to provide faster and more secure access.

Any HiPS server – having full copy and distribution rights - **may** mirror a HiPS survey from other *HiPS server* on condition that the HiPS survey is described by the *HiPS server list*, and with the explicit property "*hips_status = … clonable …*".

A HiPS server which provides a copy of a HiPS survey **must** specify "*hips_status = … mirror ...*" for the clone. If the "*hips_status*" was originally "… *clonableOnce* …" it **must** replace it by "… *unclonable* …" to avoid a new copy from this mirror. If the mirroring is partial (lower *hips_order*, or only a subset of tile formats), the server who has done the mirroring **must** specify "*hips_status = … partial …*", and adjust the relevant properties accordingly (*hips_service_url, hips_order, hips_tile_format*), in its HiPS list and in the "properties" file of the concerned HiPS.

A HiPS server which provides a copy of a HiPS survey **must not** modify the *hips_release_date* of the master site in order to detect out of date copies.

There is no dedicated protocol for the HiPS mirroring process. It can be done via wget, rsync, Hipsgen or any other method used for synchronized HTTP web sites.

# 6 HiPS access and use

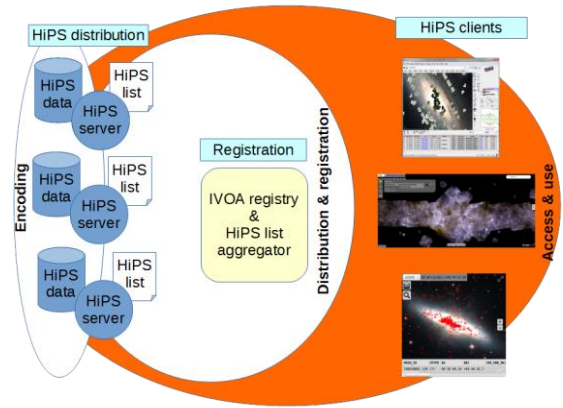

This section describes the procedures that a HiPS client **should** implement to access and use HiPS.

HiPS allows a wide range of usages. One could use the HIPS surveys in a dedicated window as part of a web application or one could use the HIPS client to find areas of interest using the tiles and then retrieve them in FITS format to gain access to their full dynamic range. With its open and simple architecture, usage is left to the imagination of the users and the capabilities of the client.

## 6.1 Local access

Some HiPS clients are capable of displaying a locally stored HiPS by using its root directory file name and deriving the file names of the various HiPS components (tiles, properties, MOC etc.). In this mode, there is no need to provide registry access or cache management. Only the knowledge of the HiPS root directory is required. Tiles and properties, and additional metadata file paths are directly derived and defined from the HiPS root directory path. Usually, the properties file is read first by the client in order to know the required HiPS parameters and limits (HiPS deepest order, tile formats, coordinate system reference, etc).

- *properties* file => **rootDir**/properties

- Tiles => **rootDir** /Norder***K***/Dir***D***/Npix***N{.ext}***
  where ***K*** is the order, ***N*** tile index, ***D***=(***N***/10000)*10000 (integer division),
  ***{.ext}*** is .fits, .jpg, .png, …

- MOC file => **rootDir** /Moc.fits

- *Allsky* files => **rootDir** /Norder***K***/Allsky***{.ext}***
  where ***K*** is the order, ***{.ext}*** is .fits, .jpg, .png, …

- etc.

*Example: the FITS tile containing the HiPS data corresponding to the HEALPix cell index 10302 at the order 6 will be available through this relative path ./Norder6/Dir10000/Npix10302.fits*

## 6.2 Remote access

Most of the HiPS clients use remote access through the Internet. In this mode, the HiPS structure is read using the HTTP protocol.  Since the remote client knows the base URL of a dedicated HiPS, it can directly query the properties, tiles and the various additional elements of the HiPS structure.

- *properties* file => **baseURL**/properties

- Tiles => **baseURL** /Norder**K**/Dir**D**/Npix**N{.ext}**
  where **K** is the order, **N** tile index, **D**=(**N**/10000)*10000 (integer division),
  **{.ext}** is .fits, .jpg, .png, …

- MOC file => **baseURL** /Moc.fits

- *Allsky* files => **baseURL** /Norder**K**/Allsky**{.ext}**
  where **K** is the order, **{.ext}** is .fits, .jpg, .png, …

- etc.

*Example: the FITS tile containing the HiPS data corresponding to the HEALPix cell index 10302 at the order 6 located on "path/of/myhips" HTTP location on "myhostname" server will be available through this URL http://myhostname/path/of/myhips/Norder6/Dir10000/Npix10302.fits*

The client knowledge of any available HiPS, their name, their base URL and possibly their mirror sites, **may** be read through the IVOA registry declarations (see 5.3).

In the case of remote HiPS access, it is strongly recommended that the client implement a cache mechanism to minimize load on the HiPS servers.

There is no recommended HiPS authentification and/or encryption mechanism. As remote HiPS access is based on any regular HTTP server, any usual HTTP/HTTPS authentification and encryption method **may** be applied if required (see IVOA SSO document [12]).

## *6.3 Suggested display algorithms*

There is no recommended HiPS client display algorithm. The best HiPS display algorithms depends on the goal of the client.

The main goal of the display algorithm is to draw the relevant tiles at the appropriate resolution for a particular sky region. The larger the sky region, the more complex the display algorithm will need to be in order to take into account the distortions due to the spherical projection whatever this projection is (sinus, tangential, aitoff, mollweide, etc…).

The two next sections will discuss two display algorithms for HIPS *images* and *catalogues* which could be adapted according to different scientific goals. These algorithms are usable for any sky region size.

### 6.3.1 Algorithm for drawing *Image HiPS*

The best HiPS display strategy is usually a balance between the drawing speed and the display quality. The following is an example of the basic algorithm, with emphasis on a number of improvements as implemented in the Aladin clients.

**Basis of the algorithm**

For a given sky region (typically a rectangle area corresponding to the client display window):

1. Compute the relevant HiPS order (one screen pixel should be cover by one tile pixel)
2. Compute the list of HEALPix tile indices covering the region (HEALPix lib function)
3. Retrieve the corresponding tiles (locally or via the internet)
4. Draw each tile on the corresponding sky location based of the associated HEALPix cell. The fast method uses only the 4 corners of the HEALPix cell and draw the tile in two complementary triangles (clipping required) mapping the bilinear stretch of the tile. This drawing step may be improved in quality by using more control points than the 4 corners by subdividing each tile in sub HEALPix cells and by this way, reduce the projection distortions.

***Allsky* usage improvement**

The client drawing a HiPS at low resolutions (order 0 to 3) should first determine whether a *Allsky.xxx* file exists in the NorderK directory, and if yes, it should load it, split it as a collection of very low resolution tiles, and draw them. A client could also decide to use the regular individual NorderK tiles as an intermediate zoom level before drawing the Norder(K+1) tiles.

***Moc.fits* usage improvement**

A client may use the Moc.fits file to avoid loading any HiPS tiles located outside the HiPS coverage (intermediate step between 2 and 3 in the previous algorithm).

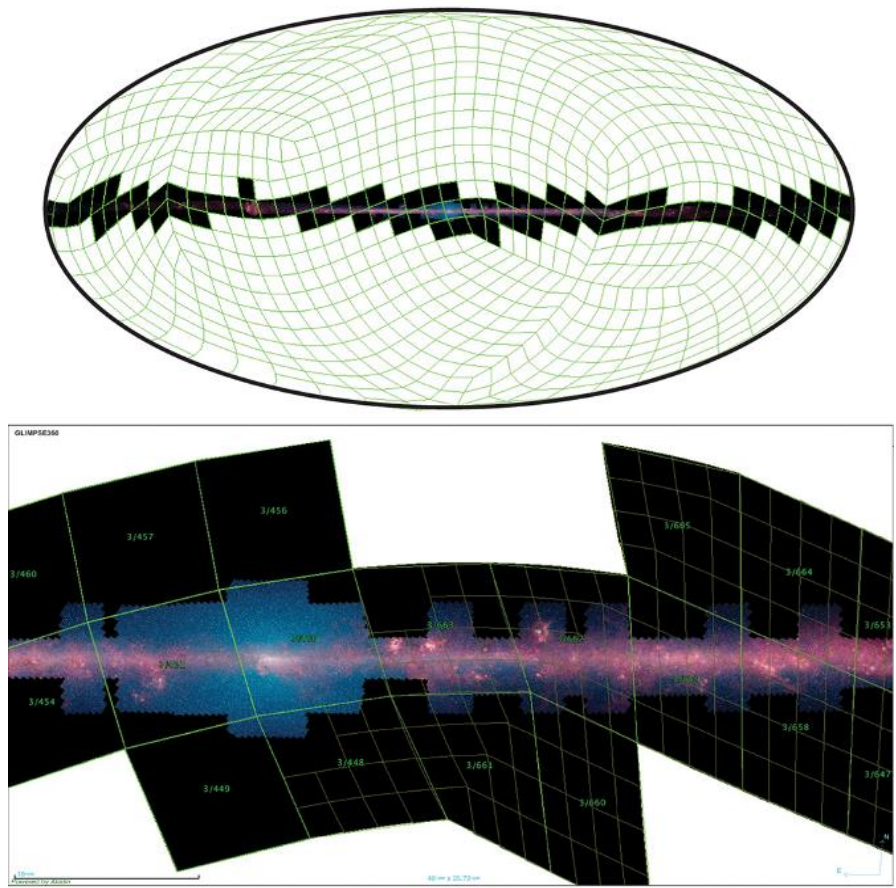

*9 - HiPS tiles of Glimpse colored survey*

### 6.3.2 Algorithm for drawing *Catalogue HiPS*

The algorithm for drawing catalogue HiPS is simpler to implement than the one for the image HiPS because the catalogue tiles are just used as a container where the sky position for of each source in the catalogue is provided explicitly.

**Basis of the algorithm**

For a given sky region:

1. From the lower order:
    1. Compute the list of HEALPix tile indices covering the region (HEALPix lib function)
    2. Retrieve the corresponding tiles (locally or via the net)
    3. Draw the sources inside the tiles (position on the sky and/or associated measurement information)
    4. *Continue with the next order if required (depending of the algorithm: number of drawn sources, order number compared to the view size…)*

**Allsky usage improvement**

The client which displays a catalogue HiPS at low resolution (order 0 to 3) may look first if a *Allsky.xxx* is existing in the NorderK, and if yes, it should load it and display the corresponding sources. In this case, the regular NorderK tiles must be ignored for avoiding duplicates sources.

**Moc.fits usage improvement**

A client may use the Moc.fits file to avoid loading the HiPS tiles outside the HiPS coverage (intermediate step between 2 and 3 in the previous algorithm).

**metadata.xml usage improvement**

A client may use the *metadata.xml* file for knowing the descriptions, the units and other related metadata associated to the catalogue HiPS.

# Appendix

This section provides a set of basic rules and formulae useful for HiPS manipulations.

- *Hierarchy path of tile*
  Tile **N** in order **K** → Norder**K**/Dir**D**/Npix**N{.ext}**
  where **D**=(**N**/10000)*10000 (integer division),
  and **{.ext}** is depending of the file format of the tiles (.fits, .jpg, .png …)

- *Parent of the tile*
  Tile **N** in order **K** → Direct parent order **K-1**: tile **N/4** (integer division)

- *Children of the tile*
  Tile **N** in order **K** → Children order **K+1**: *tiles **Nx4, Nx4+1, Nx4+2, Nx4+3***

- *Number of HEALPix cells on the whole sphere*
  Order $K \rightarrow 12 \times 2^K \times 2^K$

- *HEALPix cell angular size (global average)*
  *HEALPix cell angular size* at order $K =\sim$ sqrt( $4 \times$ PI / ($12 \times 2^K \times 2^K$))

- *Spherical coordinate of a HEALPix cell (center)*
  Cell indice **N**, order **K** → (**lon,lat**): requires a HEALPix method[5]:
  (**lon,lat**) = HEALPix_pix2ang_nested(**N,K**)

- *Spherical coordinates of a HEALPix cell (4 corners)*
  Cell indice **N**, order **K** → (**lon0,lat0,…lon3,lat3**): requires a HEALPix method:
  (**lon0,lat0,…lon3,lat3**) = HEALPix_pix2corners_nested(**N,K**)

- *HEALPix cell containing a spherical coordinate*
  (**lon,lat**), order **K** → Cell indice **N**: requires a HEALPix method:
  **N** = HEALPix_ang2pix_nested(**lon,lat,K**)

- *Image HiPS size vs original image collection size*
  HiPS size → ~ 1.3x the original image collection
  (without overlay and without hole in original image collection, using the same coding (BITPIX) and the same pixel angular resolution).

- *Image HiPS size order K vs HiPS size order K+1*
  HiPS size order **K** → ~ 0.25x HiPS size order **K+1**

- *Catalog HiPS size vs original catalog size*
  HiPS size ~= original catalog size

---

[5]    Available on the HEALPix Web site: http://healpix.sourceforge.net/